\documentclass[conference,a4paper]{IEEEtran}
\usepackage[cmex10]{amsmath} 
\usepackage{amssymb,amsthm,cite}
\usepackage[dvips]{graphics}
\usepackage{graphicx}
\usepackage{psfrag}
\usepackage{subfigure}
\usepackage{empheq}
\usepackage{mathtools}
\usepackage{algorithm}
\usepackage{algorithmic}
\usepackage{color}
\usepackage{tcolorbox}
\usepackage{dsfont}
\usepackage{float}
\usepackage{stfloats}
\usepackage{lipsum}

\DeclareMathOperator*{\argmax}{\arg\max}

\allowdisplaybreaks

\newcommand{\RNum}[1]{\uppercase\expandafter{\romannumeral #1\relax}}

\newtheorem{theorem}{Theorem}

\theoremstyle{definition}
\newtheorem{remark}{Remark}

\makeatletter
\def\blfootnote{\gdef\@thefnmark{}\@footnotetext}
\makeatother

\def\cX{{\mathcal X}}

\def\bR{{\mathbb R}}
\def\bE{{\mathbb E}}

\title{Computing the Feedback Capacity of Finite State Channels using Reinforcement Learning}
\author{
\IEEEauthorblockN{Ziv Aharoni} 
    \IEEEauthorblockA{
    Ben-Gurion University of the Negev\\
    zivah@post.bgu.ac.il }
\and
\IEEEauthorblockN{Oron Sabag}
    \IEEEauthorblockA{
    Ben-Gurion University of the Negev\\
    oronsa@post.bgu.ac.il}
\and
\IEEEauthorblockN{Haim H. Permuter} 
    \IEEEauthorblockA{
    Ben-Gurion University of the Negev \\
    haimp@bgu.ac.il }
}

\begin{document}
\maketitle

\begin{abstract}
In this paper, we propose a novel method to compute the feedback capacity of channels with memory using reinforcement learning (RL). 
In RL, one seeks to maximize cumulative rewards collected in a sequential decision-making environment. 
This is done by collecting samples of the underlying environment and using them to learn the optimal decision rule.
The main advantage of this approach is its computational efficiency, even in high dimensional problems.
Hence, RL can be used to estimate numerically the feedback capacity of unifilar finite state channels (FSCs) with large alphabet size. 
The outcome of the RL algorithm sheds light on the properties of the optimal decision rule, which in our case, is the optimal input distribution of the channel.
These insights can be converted into analytic, single-letter capacity expressions by solving corresponding lower and upper bounds.
We demonstrate the efficiency of this method by analytically solving the feedback capacity of the well-known Ising channel with a ternary alphabet.
We also provide a simple coding scheme that achieves the feedback capacity.
\end{abstract}

\section{Introduction}

\par Computing the capacity of a finite state channel (FSC) is a difficult task that has been vigorously researched in recent decades \cite{Gallager68}. 
With the presence of feedback, the feedback capacity of a FSC can be expressed using the directed information \cite{permuter2009finite,TatikondaMitter_IT09}.
Despite the fact that the directed information is a multi-letter expression, it was shown that it can be formulated as a Markov decision process (MDP), which enables its computability using known MDP algorithms \cite{PermuterCuffVanRoyWeissman08}. 

\par When formulated as a MDP, the feedback capacity of a FSC can be computed using a variety of methods, such as value and policy iteration. 
These algorithms have been proven very effective for channels with relatively small alphabets of the channel input, output and state \cite{PermuterCuffVanRoyWeissman08, Ising_channel, Sabag_BEC, POSTchannel,  Ising_artyom_IT,Sabag_BIBO_IT, trapdoor_generalized}. 
However, a principal drawback is that their computational complexity grows with the cardinality of the channel alphabet. 
Indeed, even for channel parameters from the ternary alphabet, these algorithms might be intractable. % implementation is time-consuming if not tractable. 
% In this paper, we address this drawback by suggesting a new algorithm to compute the channel capacity for large alphabets of the channel parameters.

\par We propose a machine learning (ML) approach to compute the capacity of such channels. ML has been proven to be a useful tool with a great impact in many research fields. One example in communications is \cite{deepcode}, wherein a learning-based algorithm was applied to design a reliable code for the additive white Gaussian noise channel with feedback. The present work introduces a new role of ML in communications, an efficient computation of multi-letter capacity expressions using RL algorithms. 

\par We propose a methodology that uses RL to compute the feedback capacity of unifilar FSCs.
Initially, a RL algorithm, namely the deep deterministic policy gradient (DDPG), is used to numerically estimate the feedback capacity.
Then, the outcome of the RL algorithm is used to conjecture the structure of the analytic solution, which is expressed by a directed graph.
The conjectured graph, that is called a \emph{Q-graph}, can be used to compute analytic lower and upper bounds of the feedback capacity \cite{Q-UB}.
The bounds are guaranteed to coincide to the feedback capacity, in the case that the Q-graph of the analytic solution is extracted.
Furthermore, the Q-graph can be used to derive a simple, capacity-achieving coding scheme of the channel.
In our work, the proposed methodology enabled us to compute the feedback capacity of the Ising channel with a ternary alphabet (Ising3), and derive a capacity achieving coding scheme.

\par The remainder of the paper is organized as follows.
Section \ref{sec:pre} includes the notation and preliminaries.
In Section \ref{sec:main-results}, we present our main results.
Section \ref{sec:rl} provides background on RL and on the DDPG algorithm.
In Section \ref{sec:numerical}, we estimate the feedback-capacity of the Ising3 using the DDPG algorithm.
In Section \ref{sec:analytical}, we prove the feedback-capacity of the Ising3 channel and present a simple capacity-achieving coding scheme.
Section \ref{sec:conclusions} contains conclusions and a discussion of the future work.

\section{Notation and Problem Definition} \label{sec:pre}
\subsection{Notation}
Calligraphic letters, $\mathcal{X}$, denote alphabet sets, upper-case letters, $X$, denote random variables, and lower-case letters, $x$, denote sample values.
A superscript, $x^t$, denotes the vector $(x_1,\dots,x_t)$.
The probability distribution of a random variable, $X$, is denoted by $p_X$.
We omit the subscript of the random variable when it and the argument have the same letter, e.g. $p(x|y)=p_{X|Y}(x|y)$.
The binary entropy is denoted by $H_2(\cdot)$
\subsection{Unifilar state channels}\label{app:unifilar}
A FSC is defined by the triplet $(\mathcal{X}\times\mathcal{S},p(y,s^\prime|x,s) ,\mathcal{Y}\times\mathcal{S})$, where $X$ is the channel input, $Y$ is the channel output, $S$ is the channel state at the beginning of the transmission, and $S^\prime$ is the channel state at the end of the transmission, where the cardinalities $\mathcal{X},\mathcal{Y},\mathcal{S},$ are assumed to be finite.
At each time $t$, the channel has the memory-less property
\begin{equation}\label{eqn:unifilar}
  p(s_t,y_t|x^t,s^{t-1},y^{t-1})=p(y_t|x_t,s_{t-1})p(s_t|x_t,s_{t-1},y_t).
\end{equation}
A FSC is called \textit{unifilar} if the new channel state, $s_t$, is a time-invariant function of the triplet $s_t=f(x_t,y_t,s_{t-1})$.
For a FSC with feedback, the input $x_t$ is determined by the message and the feedback tuple $y^{t-1}$.

The feedback capacity of a unifilar FSC is given by a multi-letter expression that is presented in the following theorem.
\begin{theorem}\label{thm:capacity_unifilar}\cite[Thm 1]{PermuterCuffVanRoyWeissman08}
The feedback capacity of a strongly connected unifilar state channel, where the initial state $s_0$ is available to both to the encoder and the decoder, can be expressed by
\begin{align*}
    C_{\text{fb}} &= \lim_{N\rightarrow \infty} \sup_{\{p(x_t|s_{t-1},y^{t-1})\}_{t=1}^N} \frac{1}{N} \sum_{i=1}^N   I(X_i,S_{i-1};Y_i|Y^{i-1}).
\end{align*}
\end{theorem}
\subsection{Ising3 channel}
The Ising channel model was introduced as an information theory problem by Berger and Bonomi in $1990$ \cite{Berger90IsingChannel}, 70 years after it was introduced as a problem in statistical mechanics by Lenz and his student, Ernst Ising \cite{ising1925beitrag}.
Berger and Bonomi studied the channel with a binary alphabet size.
We investigate a generalized version of the Ising channel, where the alphabets are not necessarily binary.
The Ising channel is defined by
\begin{align}
    Y &=     \begin{cases}
                X   &, \text{w.p } 0.5   \\
                S   &, \text{w.p } 0.5
            \end{cases},     \label{eqn:ising_out}\\
    S^\prime &= X. \label{eqn:ising_state}
\end{align}
Hence, if  $X=S$ then $Y=X=S$ w.p 1. Otherwise, $Y$ is assigned by one of the last two symbols with equal probability.

\section{Main Results}\label{sec:main-results}
The following theorems constitute our main results.
\begin{theorem}\label{thm:main_RL}
    The feedback capacity of a unifilar FSC can be estimated by a RL algorithm.
\end{theorem}
\begin{remark}
    Theorem \ref{thm:main_RL} is a computational result. 
    Specifically, while previous estimations of the capacity were constrained by the cardinality of the channel parameters, we show that the RL algorithm is dimensional free. 
\end{remark}
Using the numerical results from the RL algorithm, one can deduce the analytic solution structure by a Q-graph \cite{Q-UB}, which is used to compute the feedback capacity.  

\par The following theorem is an instance of a known channel that we were able to solve using the numerical results from the RL algorithm.
\begin{theorem} \label{thm:ising3_cfb}
    The feedback-capacity of the Ising3 channel is given by 
    \begin{equation}
        C_{fb} = \max_{p \in [0,1]} 2\frac{H_2(p) + 1-p}{p+3},
    \end{equation}
    where $C_{fb} \approx 0.961227$ for $p \approx 0.263805$.
\end{theorem}
Furthermore, we derive a simple coding scheme that achieves the feedback capacity in Theorem \ref{thm:ising3_cfb}.
\begin{theorem}\label{thm:code-scheme}
There exists a simple coding scheme for the Ising channel with general alphabet $\mathcal X$, with the following achievable rate:
\begin{align}
    R(\mathcal X) &= \max_{p \in [0,1]}  2\frac{H_2(p) + (1-p) \log\left(\lvert\cX\lvert-1\right)}{p+3}.
\end{align}
\end{theorem}
\noindent Note that for $|\mathcal X| = 3$, the coding scheme achieves the capacity in Theorem \ref{thm:ising3_cfb}.

\par The coding scheme is described by a repeated procedure that is given by the following:

\noindent\textbf{Code construction and initialization:}
\begin{itemize}
    \item[-] The message is a stream of $n$ uniform bits.
    \item[-] Transform the message into a stream of symbols from $\mathcal X$, denoted by $\nu_1\nu_2\dots$ with the following statistics:
    \begin{equation}
        \nu_i = \begin{cases}
                    \nu_{i-1} &, \text{w.p } p \\
                    \text{Unif}[\cX \backslash \nu_{i-1}] &, \text{w.p } 1-p
                \end{cases}
    \end{equation}
    In words, a new symbol equals the previous symbol with probability $p$ and, otherwise, it is randomly chosen from the remaining symbols. 
    This mapping can be done, for instance, by using enumerative coding, as shown in \cite{1054929}.
    \item[-] At the first time, the encoder transmits $\nu_1$ twice.
    \item[-] The decoder, upon receiving $y_1,y_2$, decode $\hat{\nu}_1 = y_2$ and sets $c=2$.
\end{itemize}
The transmission procedure is given by the following:\\
\textbf{Encoder:}
\begin{enumerate}
    \item If $\nu_t = \nu_{t-1}$ transmit $\nu_t$ twice and move to the next symbol.
    \item If $\nu_t \neq \nu_{t-1}$ transmit $\nu_t$ once and view the last feedback $y$.
    \begin{enumerate}
        \item If $y = \nu_t$ move to the next symbol.
        \item If $y \neq \nu_t$ transmit $\nu_t$ again and move to the next symbol.
    \end{enumerate}
\end{enumerate}
\textbf{Decoder:}
\begin{enumerate}
    \item If $y_t \neq \hat{\nu}_{c-1}$ then $\hat{\nu}_c = y_t$, increment $c=c+1$. 
    \item If $y_t = \hat{\nu}_{c-1}$ then wait for $y_{t+1}$, set $\hat{\nu}_c=y_{t+1}$, and increment $c=c+1$.
\end{enumerate}

In Section \ref{sec:analytical}, we prove that the coding scheme yields a zero-error code and that its maximum rate equals the feedback capacity as given in Thm. \ref{thm:ising3_cfb}. 
    
\section{Reinforcement Learning} \label{sec:rl}
\par In this section, we provide the definition of the basic RL problem setting as presented in \cite{sutton2018reinforcement} and elaborate on the DDPG algorithm.
\subsection{Background}
\par
\begin{figure}[!ht]
    \centering
    \psfrag{A}[c][][0.8]{Agent}
    \psfrag{B}[c][][0.8]{Environment}
    \psfrag{C}[c][][0.8]{D}
    \psfrag{D}[l][][0.8]{$u_t$}
    \psfrag{E}[r][][0.8]{$z_t,r_t$}
    \psfrag{F}[r][][0.8]{$z_{t-1},r_{t-1}$}
    \includegraphics[scale=0.25]{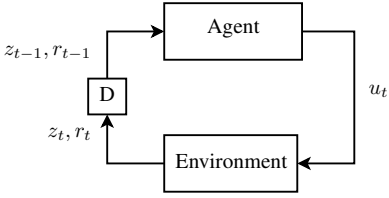}
    \caption{Depiction of the agent-environment interface in RL. The agent observes the environment state and chooses an action. In return, the environment draws an immediate reward and a next state according to $p\left(r_t \lvert z_{t-1},u_t\right),\;p\left(z_t \lvert z_{t-1},u_t\right)$. }
    \label{fig:rl_description}
\end{figure}

 The RL field in ML comprises an agent that interacts with an unknown environment by taking sequential actions. 
Formally, at time $t$, the agent observes the environment's state $z_{t-1}$ and then takes an action $u_t = A(z_{t-1})$. 
This incurs an immediate reward $r_t$ and the agent's next state $z_t$, as shown in Fig. \ref{fig:rl_description}. 

\par The environment is assumed to satisfy the Markov property,
\begin{equation} 
    p\left(z_t,r_t \lvert z^{t-1},u^t,r^{t-1}\right) = p\left(z_t,r_t \lvert z_{t-1},u_t\right). \label{eqn:markov}
\end{equation}
Hence, it can be defined by the conditional probabilities $p\left(z_t \lvert z_{t-1},u_t\right),p\left(r_t \lvert z_{t-1},u_t\right)$\footnote{One can show that the marginal probabilities are sufficient since the objective is to maximize additive rewards}.

\par The agent's policy is a sequence of actions $\pi = \{u_1, u_2, \dots \}$, and the cumulative rewards with respect to the policy from time $t$ onward are defined by $G_t = \sum_{k=t}^\infty \gamma^{k-t} r_k $, where $\gamma \in [0,1]$ is the discount factor.
The agent's goal is to find an optimal policy $\pi^\ast$  such that 
\begin{equation}
    \pi^\ast = \argmax_\pi \bE_\pi\left[ G_t \right]. \label{eqn:rl_critetia}
\end{equation}
The subscript $\pi$ of the expectation represents its dependence on the policy. 
\par In the next section, we present the state-action value function that is used as a tool to find $\pi^\ast$. 
\subsection{State-action Value function}
\par The state-action value function $Q_\pi(z,u)$  is defined as 
\begin{equation}
    Q_\pi(z,u) = \bE_\pi \left[ {G_t \lvert Z_t=z, U_t=u} \right]. \label{eqn:Q}
\end{equation}
That is, the expected cumulative rewards for taking action $u$ at state $z$ and thereafter following policy $\pi$.
Using the Markov property (Eq. \eqref{eqn:markov}) of the environment, one can decompose Eq. \eqref{eqn:Q} to
\begin{align}\
    Q_\pi(z,u) = &\bE \left[ r_t \lvert Z_t=z, U_t=u \right] + \nonumber\\  
                 &\gamma \bE_\pi \left[ Q_\pi(Z_{t+1}, U_{t+1})\lvert Z_t=z, U_t=u\right]. \label{eqn:Q_decompose}
\end{align}
The decomposition in \eqref{eqn:Q_decompose} is essential when estimating the function $Q_{\pi}(\cdot,\cdot)$ when $\pi$ is fixed.
Once the state-action value function is estimated, it forms the basis for the improvement of a given policy.
That is, for each state $z$, the current action $u(z)$ can be improved to the action $u^\prime(z)$ by choosing
\begin{equation}
    u^\prime(z) = \argmax_{u} Q_\pi(z, u).
\end{equation}
\subsection{Function approximation}

\par
\begin{figure}[h!]
    \centering
    \psfrag{A}[c][][1.]{$\substack{\text{Actor}\\\text{NN}}$}
    \psfrag{B}[c][][1.]{$\substack{\text{Critic}\\\text{NN}}$}
    \psfrag{C}[c][][0.7]{$z$}
    \psfrag{D}[c][][0.7]{$A_\mu(z)$}
    \psfrag{E}[l][][0.7]{$Q_\omega\left(z,A_\mu(z)\right)$}

    \includegraphics[scale=0.25]{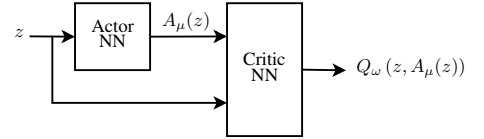}
    \caption{Depiction of actor and critic networks. The actor network comprises a NN that maps the state $z$ to an action $u=A_\mu(z)$. The critic NN maps $z, u$ pair to the estimated future cumulative rewards.}
    \label{fig:actor-critic}
\end{figure}

 The \textit{function approximators} in RL are parameterized models for $Q_\pi(z,u), A(z)$.
The \textit{actor} is defined by $A_\mu(z)$, a parametric model of $A(z)$, whose parameters are $\mu$.
The \textit{critic} is defined by $Q_\omega(z,u)$, a parametric model of the state-action value function the corresponds to $A_\mu(z)$, whose parameters are $\omega$.
Generally, the actor and critic are modeled by neural networks (NNs), as shown in Fig \ref{fig:actor-critic}.

\par The function approximations are modeled by a differentiable parametric model.
Hence, learning $Q_\pi(z,u),A(z)$ can be done without visiting the entire state space.
Specifically, the approximation interpolates its estimate for observed states to unobserved states, which enables the algorithm to converge without visiting the entire state space.
This constitutes the main difference between the RL approach and previous methods, such as DP, which turn it into a tractable solution for channels with high cardinality.

\subsection{DDPG algorithm}
The DDPG algorithm \cite{ddpg} is a deep RL algorithm for deterministic action and continuous state and action spaces. 
The training procedure comprises $N_{ep}$ episodes, where each episode contains $T$ sequential steps. 
A single step of the algorithm comprises two parallel operations: (1) collecting experience from the environment, and (2) training the actor and critic networks to obtain the optimal policy. 

\par In the first operation, the agent collects experience from the environment. 
Given the current state $z_{t-1}$, the agent chooses an action $u_t$ according to a $\epsilon$-greedy policy, where $\epsilon \in [0,1]$. 
That is, with probability $1-\epsilon$ the agent acts according to $A_\mu(z_{t-1})$, and with probability $\epsilon$ the agent takes a random action uniformly over the action space. 
The term $\epsilon$ denotes the exploration parameter, and it is crucial to encourage the agent to search the entire state and action spaces. 
Then the agent samples from the environment the incurred reward $r_t$ and the next state $z_t$. 
The transition tuple $\left(z_{t-1},u_t,r_t,z_t\right)$ is then stored in a \textit{replay buffer}, a bank of experience, that is used to improve the actor and critic networks.
Finally, the agent updates its current state to be $z_t$ and moves to the next step. 

\par The second operation entails training the actor and critic networks.
First, $N_{mb}$ transitions are drawn randomly from the replay buffer. 
Second, for each transition, we compute its target based on the right-hand side of Eq. \ref{eqn:Q_decompose}.
\begin{equation}
    y_i = r_i  + \gamma Q_\omega\left(z^\prime_i, A_\mu(z^\prime_i)\right), \quad \forall i=1,\dots,N_{mb}.
\end{equation}
Then we minimize the following objective with respect to the parameters of the critic network $\omega$ as given by
\begin{equation}\label{eqn:critic-update}
    L\left( \omega \right) = \frac{1}{N_{mb}} \sum_{i=1}^{N_{mb}} \left[Q_\omega\left(z_i, A_\mu(z_i)\right) - y_i \right]^2.
\end{equation}
The aim of this update is to train the Critic to comply with Eq. \eqref{eqn:Q_decompose}.
Afterward, we train the actor to maximize the critic's estimation of future cumulative rewards. 
That is, we train the actor to choose actions that result in high cumulative rewards according to the critic's estimation.
The actor update formula is given by
\begin{equation} \label{eqn:actor-update}
    \nabla_\mu Q_\omega(z,A_\mu(z)) =  \frac{1}{N_{mb}} \sum_{i=1}^{N_{mb}} \nabla_a Q_\omega\left(z_i, a\right)\lvert_{a=A(z_i)} \nabla_\mu A\left( z_i\right).
\end{equation}

\par
\begin{figure}[!ht]
    \centering
    \psfrag{A}[c][][0.7]{$z_{t-1}$}
    \psfrag{B}[c][][1.0]{$\substack{\text{Actor}\\\text{NN}}$}        
    \psfrag{C}[c][][0.7]{$A_\mu(z_{t-1})$}
    \psfrag{D}[c][][0.7]{$\epsilon$-greedy}        
    \psfrag{E}[c][][0.7]{$u_t$}
    \psfrag{F}[c][][0.7]{environment}        
    \psfrag{G}[l][][.7]{$z_{t-1},u_t,r_t,z_t$}
    \psfrag{H}[c][][.7]{$z_t$}
    \psfrag{J}[c][][0.7]{D}        
    \psfrag{K}[c][][0.7]{replay-buffer}        

    \psfrag{L}[c][][0.7]{$N_{mb}$ examples}
    \psfrag{M}[c][][1.0]{$\substack{\text{Compute}\\\text{networks}\\\text{updates}}$}
    \psfrag{N}[c][][0.7]{update critic}        
    \psfrag{O}[c][][0.7]{update actor}        
    \psfrag{P}[c][][1.0]{$\substack{\text{Critic}\\\text{NN}}$}  
    \includegraphics[scale=0.25]{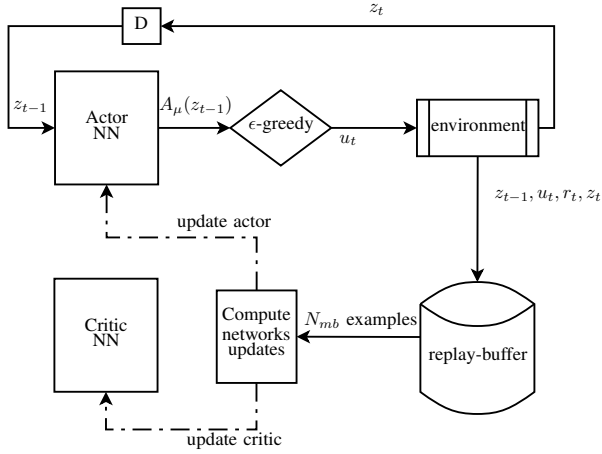}
    \caption{Depiction of the work flow of the DDPG algorithm. At each time step $t$, the agent samples a transition from the environment using $\epsilon$-greedy policy and stores the transition in the replay buffer. Simultaneously, $N_{mb}$ past transition are drawn from the replay buffer and used to update the critic and actor NN according to Equations \eqref{eqn:critic-update} and \eqref{eqn:actor-update} respectively. }
    \label{fig:ddpg-workflow}
\end{figure}

 To conclude, the algorithm alternates between improving the critic's estimation of future cumulative rewards and training the actor to choose actions that maximize the critic's estimation.
The algorithm work flow is depicted in Fig. \ref{fig:ddpg-workflow}.

\section{Estimating the capacity of the Ising3 channel using RL}\label{sec:numerical}
In This section, we show the formulation of the feedback capacity as a RL problem, including details of the implementation of the RL algorithm and the experiments we conducted on various unifilar FSCs with feedback.
\subsection{Formulation of the feedback capacity as a RL problem}
\par We formulate the Ising3 feedback capacity as a RL problem that is based on the formulation as done in \cite{PermuterCuffVanRoyWeissman08}.
We define the state by a two-dimensional vector, $z_t = \left[p\left( s_t=0 \lvert y^t\right), p\left( s_t=1 \lvert y^t\right)\right]^T$. 
The action is defined by $u_t = p\left( x_t \lvert z_{t-1}\right) \in \bR^{3\times3}$. 
The reward is defined by $r_t = I(X_t,S_{t-1};Y_t|Y^{t-1}=y^{t-1})$, which is a deterministic function of $p\left( x_t, s_{t-1}, y_t \lvert t^{t-1} \right) = z_{t-1} u_t p\left( y_t \lvert x_t,s_{t-1}\right)$. 
Hence, the conditional distribution $p\left(r_t \lvert z_{t-1},u_t\right)$ is induced by the channel distribution Eq. \eqref{eqn:ising_out}.
The next state distribution is given by the BCJR equation as given in Eq. (35) in \cite{PermuterCuffVanRoyWeissman08}. 
Accordingly, the conditional distribution $p\left(z_t \lvert z_{t-1},u_t\right)$ is induced by the channel distribution, Eq. \eqref{eqn:ising_out} and the state evolution, Eq. \eqref{eqn:ising_state}.
\subsection{Implementation of the RL algorithm}
\par We model $Q_\pi(z,u)$, $A_\mu(z)$ with two NNs, each of which is composed of three fully connected hidden layers of 300 units separated by a batch normalization layer.
The actor network input is the state $z$ and its output is a matrix $A_\mu(z) \in \bR^{3\times3}$ such that $A_\mu(z)^T \mathbf{1} = \mathbf{1}$. 
The critic network input is the tuple $\left\{ z, A_\mu(z)\right\}$ and its output is a scalar, which is the estimate for the cumulative future rewards.
In our experiments, we trained the networks for $N_{ep}=10^4$ episodes. 
Each episode length is $T=500$ steps. 
For the exploration, we chose $\epsilon=0.1$ and decayed it by $0.999$ each episode.
\subsection{Experiments}
\par We conducted several experiments to verify the effectiveness of our formulation.
First, we focused on experimenting channels whose analytic solution was proven in the past, such as the Trapdoor channel \cite{PermuterCuffVanRoyWeissman08}, Ising channel with a binary alphabet \cite{Ising_channel, Ising_artyom_IT}, Binary Erasure channel with input constraint \cite{Sabag_BEC},  and the Dicode channel \cite{Q-UB}. 
The results showed that the obtained achievable rates were within $99.99\%$ of the feedback capacity for all channels. 

\par
\begin{figure}[!ht]
    \centering
    \psfrag{A}[c][][1.1]{$Q_1$}
    \psfrag{B}[c][][1.1]{$Q_2$}
    \psfrag{C}[c][][1.1]{$Q_3$}
    \psfrag{D}[c][][1.1]{$Q_4$}
    \psfrag{E}[c][][1.1]{$Q_5$}
    \psfrag{F}[c][][1.1]{$Q_6$}
    \includegraphics[scale=0.18]{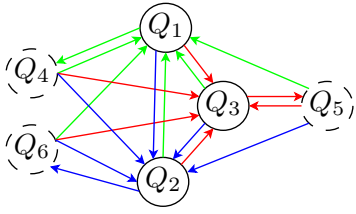}
    \caption{Q-graph showing the transitions between states as a function of the channel's output. Blue, red and green lines correspond to $Y=0,1,2$, respectively. States with dashed lines and states with solid line behave similarly.}
    \label{fig:q-graph}
\end{figure}

 Our aim is to solve a channel with large cardinality that previous methods have failed to solve due to computational complexity. 
We chose the Ising3 channel as a candidate and used our formulation to estimate its feedback capacity. 
We ran a simulation over the Ising3 channel with the same RL model as used in the previous experiments. 
By the end of training, we obtained a policy whose achievable rate is $0.96110$. 
\begin{figure}[!ht]
    \centering
    \includegraphics[scale=0.2]{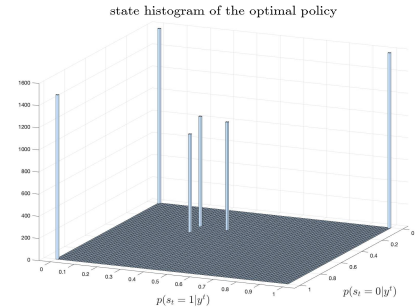}
    \caption{State histogram of the optimal policy as obtained from RL. The histogram was generated by a Monte-Carlo evaluation of the estimated policy.}
    \label{fig:state-hist}
\end{figure}
Another property of the obtained policy is that it visits only six discrete states as shown in Fig. \ref{fig:state-hist}. 
Furthermore, the transition between states is determined uniquely given the output of the channel. 
These transitions can be shown as a Q-graph, as depicted in Fig. \ref{fig:q-graph}.
\par In the next section we use the Q-graph we obtained from the estimated policy of the RL algorithm to solve the Ising3 channel.
  
\section{Analytic Solution for the Ising3 Channel}\label{sec:analytical}
In this section, we prove Theorem \ref{thm:ising3_cfb} concerning the feedback capacity of the Ising3. Specifically, we use the graphical structure in Fig. \ref{fig:q-graph} to compute a tight upper bound, and analyze the rate of the proposed coding scheme. 
\subsubsection{Bounds on the feedback capacity}
The Q-graph method, introduced in \cite{Q-UB}, is a general technique that exploits the discrete histogram in Fig. \ref{fig:state-hist} to provide upper and lower bounds on the capacity. 
The upper bound states that for any choice of a $Q$-graph,
\begin{equation} \label{eqn:Q-ub}
    C_{fb} \leq \sup_{p\left(x\lvert s,q\right)} I \left( X,S;Y \lvert Q\right),
\end{equation}
where the joint distribution is $P_{S,Q}P_{X\lvert S,Q}P_{Y\lvert X,S}$ and $P_{S,Q}$ is a stationary distribution. 
The upper-bound is tight, that is, equals to the feedback capacity, when the maximizer of \eqref{eqn:Q-ub} satisfies the Markov chain $S^\prime-Q^\prime-(Q,Y)$.

We use convex optimization tool to compute the upper-bound in  \eqref{eqn:Q-ub} with respect to the $Q$-graph in Fig. \ref{fig:q-graph}.
The result is used to conjecture a parameterized input $P_{X|S,Q}$ as the optimal solution. 
Then, using the convexity of the upper bound (as a function of the entire joint distribution), one can show that the conjectured solution is optimal, and that the upper-bound can be simplified to the expression in Theorem \ref{thm:ising3_cfb}. The tightness of the upper bound is shown via the Markov chain above.

\subsubsection{Coding scheme - Sketch of proof for Theorem \ref{thm:code-scheme}}
The coding scheme in Section \ref{thm:code-scheme} is a generalization of the optimal coding scheme for $|\mathcal X| =2$  that was presented in \cite{Ising_channel}. We analyze the achievable rate by computing the entropy rate of input symbols, divided by the expected time until decoding a single symbol.

The entropy rate can be computed from the the symbols transition entropy:
\begin{align} \label{eqn:input-len}
    H(\nu_i|\nu_{i-1}) = H_2(p) + (1-p) \log\left[\lvert\cX\lvert-1\right].
\end{align} 
The expected time until decoding a single symbol $\nu_i$ is
\begin{align}\label{eqn:code-len}
    \bE \left[ L\right] = p \cdot 2 + (1-p) \cdot 1.5.
\end{align}
That is since when $\nu_i = \nu_{i-1}$, the symbol is sent twice, and when $\nu_i \neq \nu_{i-1}$, the symbol is sent once or twice with equal probability.
The proof is completed by dividing Eq. \eqref{eqn:input-len} by Eq. \eqref{eqn:code-len} and taking a maximum over $p$.

\section{Conclusions} \label{sec:conclusions}
\par We derived an estimation algorithm of the feedback capacity of a unifilar FSC using RL.
The RL approach addresses the cardinality constraint and establishes RL as a useful tool for channels with high cardinality.
We provided an example over the Ising3 channel, where we used the insights provided by the numerical results to analytically compute its feedback capacity.
Furthermore, we showed a simple capacity-achieving coding scheme for the Ising3 channel with feedback.

\par Additionally, our preliminary results imply that we are able to solve the Ising channel for any alphabet size.
Then, we plan to solve different channels numerically and, hopefully, establish methods to induce their analytic solution, and capacity-achieving coding schemes. 
Furthermore, we plan to use the feedback capacity problem as a framework to improve the RL algorithms.

\bibliography{ref}
\bibliographystyle{IEEEtran}
\end{document}